\documentstyle[11pt,epsfig,rotating]{article}

\newcommand{\stotgp}{\mbox{$\sigma_{tot}^{\gamma{p}}$}}
\newcommand{\sigND}{\mbox{$\sigma_{ND}$}}
\newcommand{\sigGD}{\mbox{$\sigma_{GD}$}}

\newcommand{\sigDD}{\mbox{$\sigma_{DD}$}}
\newcommand{\sigI}{\mbox{$\sigma_{i}$}}
\newcommand{\GeV}{\mbox{\rm GeV}}
\newcommand{\GeVa}{\mbox{\rm GeV/c}}
\newcommand{\GeVb}{\mbox{\rm GeV/c$^2$}}
\newcommand{\MeV}{\mbox{\rm MeV}}
\newcommand{\MeVa}{\mbox{\rm MeV/c}}

\newcommand{\bh}{$ep \rightarrow e\gamma{p}$}
\newcommand{\AETBH}{$A^{BH}(y)$}

\newcommand{\gsim}{\raisebox{-0.5mm}{$\stackrel{>}{\scriptstyle{\sim}}$}}
\newlength{\dinwidth} \newlength{\dinmargin} 
\begin{document}
\setlength{\unitlength}{1mm}
\begin{titlepage}
{\tt DESY 95-162   } \\
\vspace{1cm}
\vspace*{4.cm}
\begin{center} \begin{Large}

{\bf Measurement of the Total Photon Proton Cross Section \\
      and its Decomposition \\
      at 200~GeV Centre of Mass Energy} \\

\vspace{1.cm}
{H1 Collaboration}    \\
\end{Large}
\vspace*{4.cm}
{\bf Abstract:}
\end{center}
\begin{quotation}
\renewcommand{\baselinestretch}{1.0}\large\normalsize

 We present a new measurement of the total photoproduction
 cross section performed with the H1 detector at HERA.
 For an average centre of mass energy of 200~GeV a value
 of $\sigma_{tot}^{\gamma{p}}= 165\pm2\pm11\mu$b has been obtained.
 A detailed analysis of the data in adequate kinematic regions
 enabled a decomposition of the total cross section
 in its elastic, single diffractive dissociation
 and remaining non-diffractive parts, based on
 safe assumptions on the double diffractive dissociation contribution.

\renewcommand{\baselinestretch}{1.2}\large\normalsize

\end{quotation}
\end{titlepage}
\begin{Large} \begin{center} H1 Collaboration \end{center} \end{Large}
\begin{flushleft}
 S.~Aid$^{14}$,                   
 V.~Andreev$^{26}$,               
 B.~Andrieu$^{29}$,               
 R.-D.~Appuhn$^{12}$,             
 M.~Arpagaus$^{37}$,              
 A.~Babaev$^{25}$,                
 J.~B\"ahr$^{36}$,                
 J.~B\'an$^{18}$,                 
 Y.~Ban$^{28}$,                   
 P.~Baranov$^{26}$
% ...
\end{flushleft}
\begin{flushleft} {\it
% ...
$ ^{12}$ DESY, Hamburg, Germany$^a$ \\
% ...
$ ^{14}$ II. Institut f\"ur Experimentalphysik, Universit\"at Hamburg,
                                     Hamburg, Germany$^ a$  \\
% ...
$ ^{18}$ Institute of Experimental Physics, Slovak Academy of
                Sciences, Ko\v{s}ice, Slovak Republic$^ f$\\
% ...
$ ^{25}$ Institute for Theoretical and Experimental Physics,
                                                 Moscow, Russia \\
 $ ^{26}$ Lebedev Physical Institute, Moscow, Russia$^ f$ \\
% ...
$ ^{28}$ LAL, Universit\'{e} de Paris-Sud, IN2P3-CNRS,
                            Orsay, France\\
 $ ^{29}$ LPNHE, Ecole Polytechnique, IN2P3-CNRS,
                             Palaiseau, France \\
% ...
$ ^{36}$ DESY, Institut f\"ur Hochenergiephysik,
                              Zeuthen, Germany$^ a$\\
 $ ^{37}$ Institut f\"ur Teilchenphysik,
          ETH, Z\"urich, Switzerland$^ i$\\
% ...
$ ^a$ Supported by the Bundesministerium f\"ur
                                  Forschung und Technologie, FRG
 under contract numbers 6AC17P, 6AC47P, 6DO57I, 6HH17P, 6HH27I, 6HD17I,
 6HD27I, 6KI17P, 6MP17I, and 6WT87P \\
% ...
$ ^f$ Supported by the Deutsche Forschungsgemeinschaft\\
% ...
$ ^i$ Supported by the Swiss National Science Foundation\\
   } \end{flushleft}

%
\newpage
\section{Introduction}
\noindent
  The total cross section is an important quantity related to the
fundamental properties of particle interactions.
  Although measurements are available for hadron hadron and
real photon hadron collisions at low energy~\cite{pdg}, so far only the
data from $p\bar{p}$-colliders provide precise information
on the rise of the total cross section at high energy.
The $ep$ collider HERA, with $e$ and $p$ energies of $27.6$ and
$820$ GeV, provides a new source of
information on high energy photon proton collisions.

  The interaction of electrons and protons at the HERA collider is
dominated by photoproduction processes, in which the
electron scatters through small angles
emitting a quasi-real photon, which then interacts with the proton.
  Recently the total photoproduction cross section has been measured
at HERA at a $\gamma{p}$ CMS energy of $195$~GeV by H1~\cite{stot_h1}
$\stotgp=156\pm{18}~\mu$b~\footnote
 {Note that this number underestimates $\stotgp$ by 6--7\% due to
the approximation used in the theoretical expression
   for the photon flux.}
and at a $\gamma{p}$ CMS energy of $180$~GeV by ZEUS~\cite{stot_zeus}
$\stotgp=143\pm{17}~\mu$b.
  These measurements confirmed the expected rise of the total $\gamma{p}$
cross section with energy.
  However, large systematic
uncertainties do not yet allow discrimination between different models
predicting a moderate rise of the $\gamma{p}$ total cross
section~[4-8].
  In the previous H1 analysis~\cite{stot_h1},
the systematic error is dominated by
the assumptions on the partial $\gamma{p}$ cross sections,
which is important as they have different acceptance.
  ZEUS~\cite{stot_zeus} determined the
fraction of $\gamma{p}$ diffractive events directly from the data
and thus reduced the model dependence of the result. However, the precision
of this measurement is limited by the large systematic error in the
efficiency of tagging the scattered electrons.

  In this paper a new determination of $\stotgp$ at the
average centre of mass energy
$W_{\gamma{p}}=200~\GeV$ is presented.
  Dedicated trigger conditions, specially designed for high energy
 photoproduction, allow the measurement of the diffractive components
 of the $\gamma{p}$
cross sections and, therefore, substantially reduce the model dependence
in the acceptance calculations.
  This, together with a better understanding of the electron tagging
efficiency improves the accuracy of the $\stotgp$ measurement
compared with the earlier results from HERA.

   The data used in the present analysis were taken
during a period in which HERA was operated with a
positron beam. Nevertheless,
``electron" is used as generic term for the HERA beam lepton
throughout this paper.

\section{Photoproduction}

In $ep$ collisions the total
photoproduction cross section, $\stotgp$ can be related to
the total differential
$ep$ cross section by the Weizs\"acker-Williams formula~\cite{wwa}
for the photon flux~$F(y,Q^2)$
\begin{eqnarray*}
  \frac{d^2\sigma^{ep}(s)}{dydQ^2} =
  \stotgp{(ys)}\cdot(1+\delta_{RC})\cdot F(y,Q^2)=
  \hspace{6.0cm}
\end  {eqnarray*}
  \vspace{-0.7cm}
\begin{eqnarray}
  \hspace{2.0cm}
  =\stotgp{(ys)}\cdot(1+\delta_{RC})\cdot \frac{\alpha}{2\pi{Q^2}}
  \Bigl(\frac{1+(1-y)^2}{y}
  -\frac{2(1-y)}{y}
  \cdot\frac{Q^2_{min}}{Q^2}\Bigr),
  \label{eqWWA}
\end  {eqnarray}
where $Q^2$ is the negative square of the photon 4-momentum, or the
virtuality of the photon, and $s$ is the squared
centre of mass energy of the $ep$ interaction.
For small scattering
angles, $y$ is defined as
$1-{E'}_e/{E}_e$, where ${E}_e$ and ${E'}_e$ are the energies
of the initial and scattered electron respectively.
The  minimum photon virtuality is $Q^2_{min}=(m_ey)^2/(1-y)$.
The factor $(1+\delta_{RC})$
takes into account QED radiative corrections to the $ep$ Born cross
section.
  In the formula~(\ref{eqWWA}) a dependence of $\stotgp$ on $Q^2$ and
a contribution of longitudinally
polarized photons are neglected. These are good approximations in
the present kinematic conditions~\cite{lebedev}.

  Hadronic final states produced in real photon proton collisions
resemble those observed in hadron hadron collisions.
  This similarity led to the phenomenological approach
to describe photoproduction by the so called Vector Meson Dominance~(VMD)
model~\cite{vmd}, where
the photon first converts into a vector meson (predominantly
the $\rho^0$) which then interacts with the proton.
  As in hadron hadron collisions, the total $\gamma{p}$ cross section
has a substantial contribution from diffractive $\gamma{p}$ reactions,
which have a final state topology radically different
from the bulk of non-diffractive events. Since
  diffractive reactions involve no exchange of quantum numbers between
the incident particles, the final state is
characterized by the appearance of large rapidity intervals, or gaps,
with no hadrons.
  This feature of diffractive events is exploited below
to determine their contributions to the $\gamma{p}$ cross section.
 We distinguish the following diffractive processes in
photoproduction:
\begin{itemize}
\item Elastic vector meson production~(EL)
  $ \gamma + p \rightarrow V + p$, where $V$ stands for one of the
  vector mesons $\rho^0, \omega, \phi$. The true electro-magnetic
  elastic
  reaction $ \gamma + p \rightarrow \gamma + p$ has a very low cross section
  and is neglected~\cite{vmd}.
\item Single photon diffractive dissociation~(GD)
  $ \gamma + p \rightarrow X + p$, where the photon dissociates
  into the heavy hadronic state~$X$ and the proton stays intact.
\item Single proton diffractive dissociation~(PD)
  $ \gamma + p \rightarrow V + Y$, where the proton dissociates into
  a hadronic state Y and a vector meson
  is produced in the photon direction.
\item Double  diffractive dissociation~(DD)
  $ \gamma + p \rightarrow X + Y$, where both the photon and the proton
  dissociate.
\end{itemize}
  For the last three diffractive reactions the cross section is considered
for the full momentum transfer range and for
masses~($M$) of the dissociating system obeying
$M^2\,<\,0.1W^2_{\gamma{p}}$, where $W_{\gamma{p}}$ is the centre of mass
energy.
  The value $0.1$ is chosen to enable direct comparisons with
measurements of the diffractive cross section from hadron colliders
and fixed target photoproduction experiments as well as with
available theoretical calculations.

All processes $\gamma + p \rightarrow X$ not belonging to the
contributions defined above are called non-diffractive~(ND).
 These processes dominantly
involve exchange of quantum numbers between the photon
 and the proton.

\section{Monte Carlo Models for Photoproduction}

  Two Monte Carlo~(MC)
models, based on the event generators PYTHIA~\cite{pythia}
 and PHOJET~\cite{phojet},
are used for the acceptance calculation.
  Both models include all the diffractive and non-diffractive
contributions to the $\gamma{p}$ cross section discussed above.

The model for non-diffractive events in the PYTHIA program is
similar to the multiple-interaction model developed for hadron hadron
collisions~\cite{TS_MvZ}. An eikonal approach is used, in which the
rate of jets above a transverse
momentum~$p_{\perp\mathrm{min}}$ (the default value
$p_{\perp\mathrm{min}}=1.45~\GeVa$ is taken)
is combined with a parameterization of the non-diffractive total cross
section to calculate a probability distribution in the number of
semi-hard interactions. For events below this cut-off
 two longitudinal strings are stretched between the proton
and the ``VMD-photon", to give a representation  of an event structure
caused by soft gluon exchange.

  Events with elastic, diffractive single and double dissociation
scattering in the PYTHIA MC model have the same general structure:
the $t$-dependence is given by a function $\exp(Bt)$,
 where $t$ is the square of the four-momentum
transfer in the diffractive reaction and
$B$ is the nuclear slope parameter. Within the energy range used
in this analysis the nuclear slope for elastic vector meson
production is $B\approx\,11~(\GeVa)^{-2}$.
A $\rho^0$ formed by $\gamma \to \rho^0$ in elastic or diffractive
scattering is transversely polarized and therefore its
decay angular distribution in $\rho^0 \to \pi^+ \pi^-$ is taken
to be proportional to $\sin^2 \theta$, where the reference axis is
given by the $\rho^0$ direction of motion.
The relative rates of $\rho^0, \omega, \phi$ production are assumed
to be about $13:1.5:1$~\cite{vmd}.

In single diffractive dissociation, the
$B$-slope is assumed to be half that of elastic scattering.
 The events are generated according to a $dM^2/M^2$ distribution for the
dissociation system of mass~$M$.
The mass spectrum of the system is assumed to start
at 0.2 \GeVb\ above the mass $M_{in}$ of the incoming particle
(using the $\rho^0$ mass for the incoming $\gamma$). A light dissociated
system,
with a mass less than 1 GeV above the mass of the incoming particle,
is taken to decay isotropically into a two-body state.
Single-resonance states, such as $N^*$
or $\omega(1600)$, are
not generated explicitly, but are described in this
average manner.
A more massive system is treated
as a string stretched along the $\gamma{p}$ interaction axis.
The secondary hadrons from the string decay are distributed in a longitudinal
phase space with limited transverse momentum.

In the event generator PHOJET, the multi-particle
non-diffractive final
states are constructed from a parameterization of the photon proton
scattering amplitude in an eikonal approximation using the two-component
Dual Parton Model~\cite{dpm}.  The coding of the model is similar
to that of the MC generator DTUJET~\cite{dtujet} simulating particle
production in $pp$ and $\bar{p}p$ collisions up to very high energies.

In the generator PHOJET elastic vector meson production is
similar to that in the PYTHIA model.
For diffractive dissociation the PHOJET model assumes a
mass dependent nuclear slope~$B$~\cite{Engel}.
This slope parameterization gives a steady transition from elastic
scattering to single and double diffractive dissociation.
The mass spectrum is generated according to a
$dM^2/(M^2-M_{in}^2)$ distribution
starting from two pion masses above the mass $M_{in}$
(using the $\rho^0$ mass for the incoming $\gamma$).
The low-mass resonance structure is taken into account
in an approximate way
to provide a phenomenological description of the dissociated
mass spectrum observed experimentally~\cite{chapin,goulianos}.
To take the transverse polarization of the
incoming photon into account, the decay of the elastically produced
vector meson resonances into two or three
particles is performed in
the $s$-channel helicity frame according to the angular
distributions given in~\cite{vmd}.
In addition to resonances, a continuous multi-particle final
state in diffraction is generated by simulating a pomeron-proton or
pomeron-photon scattering exactly as in the
Dual Parton Model used for photon hadron scattering.
The pomeron is treated like a virtual meson.
The soft and hard scatterings in diffraction are generated
according to cross sections given by Regge-parameterizations
and the QCD Parton Model, respectively.

    The generated events are fed into the H1 detector simulation
    program and are subject to the same reconstruction and analysis
    chain as the real data.

The QED radiative corrections are calculated using the HERACLES MC
program~\cite{heracles}. This takes into account
single photon emission from the lepton line as well as
the self energy correction to the Born photoproduction cross section.

\section{H1 Detector}

\begin{figure}[htb]  \centering
  \boldmath
  \begin{picture}(160,70)
  \put(5,10){\line(1,0){16}}
  \put(5,9){\line(0,1){2}} \put(21,9){\line(0,1){2}}
  \put(10,12){\large $1$ {\rm m}}
    \epsfig
    {file=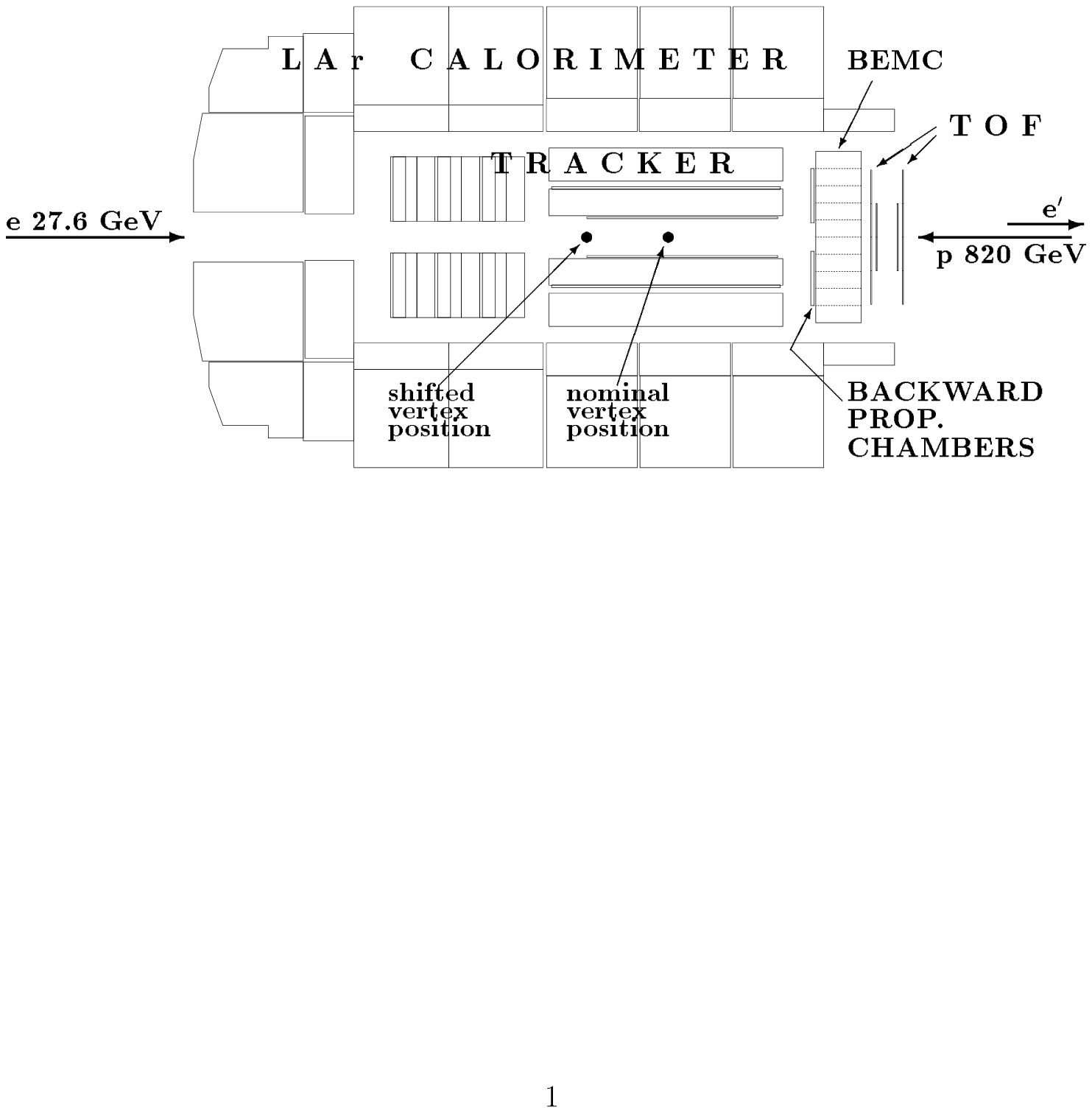,width=160mm,
      bbllx=77pt,bblly=308pt,bburx=511pt,bbury=497pt,clip= }
  \end{picture}
  \unboldmath
  \caption
  {
    A layout of the central part of the H1 detector.
    }
  \label{figH1}
\end{figure}

A detailed description of the H1 apparatus can be found elsewhere~\cite{H1NIM}.
 A schematic layout of the central H1 detector components is shown in
Fig.~\ref{figH1}.
In the following we briefly describe the components of the detector relevant
for  this analysis.

%
%
%
%

   Measurements of charged particle tracks and the interaction vertex
are provided by central
and forward tracking systems, both consisting of
drift and multi-wire proportional chambers~(MWPC).
   The central and forward track chambers cover the complete azimuthal range
and $-2.0<\eta<3.0$ in pseudo-rapidity $\eta=-\ln(\tan \frac{\theta}{2})$.
   Here $\theta$ is the polar angle with respect to the proton beam
direction (positive $z$ axis).
   The central jet chamber~(CJC) is interleaved with inner and outer double
layers of MWPC, which were used in the trigger to select
events with charged tracks pointing to the interaction region.
   This MWPC system covers the range $-1.5<\eta<1.5$.
A backward proportional chamber (BPC), with an
acceptance of $-3.0<\eta<-1.5$ allows
efficient detection of charged particles produced at large
$\theta$.

   The tracking region is surrounded by a fine grained liquid argon (LAr)
calorimeter~\cite{lar} consisting of an electro-magnetic
and a hadronic section.
   The total depth of the LAr~calorimeter varies between 4.5 and 8
hadronic interaction lengths.
   Under test beam conditions it has an energy resolution
$\sigma/E\approx 12\%/\sqrt{E/\GeV}\oplus{0.01}$ for electrons and
$\approx 50\%/\sqrt{E/\GeV}\oplus{0.02}$ for pions.
   The LAr~calorimeter covers the complete
azimuthal range and $-1.5<\eta<3.3$.
   The calorimeter is surrounded by a super-conducting solenoid providing a
uniform magnetic field of $1.15$ T parallel to the beam axis in the tracking
region.

   The time of flight system~(ToF) is located
at $z\,\approx\,-2\,${\rm m} behind
the Backward Electro-Magnetic Calorimeter~(BEMC),
which is about one hadronic interaction length deep.
   ToF is a hodoscope consisting of two planes of plastic scintillators
mounted perpendicular to the beam direction.
   The angular coverage of the ToF counters corresponds to $-3.5<\eta<-2$.
   Having a time resolution better than $2\,${\rm ns} the ToF system
enables efficient separation of $ep$ interaction events
from the upstream background.
   In the present analysis the ToF system is included in the
trigger for photoproduction events.
   The efficiency of the ToF counters has been measured using muons in
the proton beam halo and corresponds to $(98\pm{1})\%$ for minimum
ionizing particles.

   The luminosity system measuring the reaction
$ep\,\rightarrow\,e\gamma{p}$
consists of two TlCl/TlBr crystal calorimeters.
   The small angle electron detector~(electron tagger)
is located at $z=-33\,${\rm m}
and is also used to trigger on photoproduction events.
   Its $7\times{7}$ crystal matrix (an individual crystal
measures $2.2\times{2.2}${\rm cm}) accepts electrons
with an energy between  $0.2E_e$ and $0.8E_e$ and scattering
angles $\theta'\le 5\,${\rm mrad} $(\theta' = \pi -\theta)$,
corresponding to $Q^2 < Q^2_{max} =  0.01~\GeV^2$.
  The photon detector is located at $z=-103\,${\rm m} and consists of a
$5\times{5}$ crystal matrix.
  Both calorimeters are 22 radiation length deep and their energy
resolution in the present data taking period was measured to be
$\sigma(E)/E=0.15/\sqrt{E/\GeV}\oplus{0.01}$.

\section{Luminosity Measurement and Tagging Efficiency}
%
  The basic requirement used to tag quasi-real photoproduction
  processes in H1 is the detection of the scattered electron in the
  electron tagger. This guarantees very low $Q^2 < 10^{-2}$~GeV$^2$.
  Three main ingredients contribute to the overall
  precision of the measurement of the cross section \stotgp:
  the luminosity measurement error, the knowledge of the electron tagger
  acceptance and the efficiency of the main apparatus
  for triggering and reconstruction of the hadronic
  final states produced in photon proton collisions.
  The first two are discussed in this section.

%
%
\begin{figure}[b] \centering
\begin{picture}(170,110)(05,00)
\boldmath
  \put(070,095){\bf a)}  \put(149,095){\bf b)}
  \put(070,044){\bf c)}  \put(151,044){\bf d)}
  \put(058,004){\bf $E'_e$(GeV)}
  \put(060,056){\bf E$_{\gamma}$(GeV)}
  \put(143,056){\bf x$_{ET}$(cm)}
  \put(154,004){\bf y$_{vis}$}
  \put(002,060){\begin{sideways} {\small \bf
             1/N$_{\rm ev}$dN/dE$_{\gamma}$(GeV$^{-1}$)} \end{sideways}}
  \put(085,018){\begin{sideways} {\small \bf
                1/N$_{\rm ev}$dN/dy} \end{sideways}}
  \put(002,014){\begin{sideways} {\small \bf
                d$\sigma$/dE (mb/GeV)} \end{sideways}}
  \put(085,067){\begin{sideways} {\small \bf
                $<$E$_{ET}\!\!>$ (GeV)} \end{sideways}}
 \epsfig
    {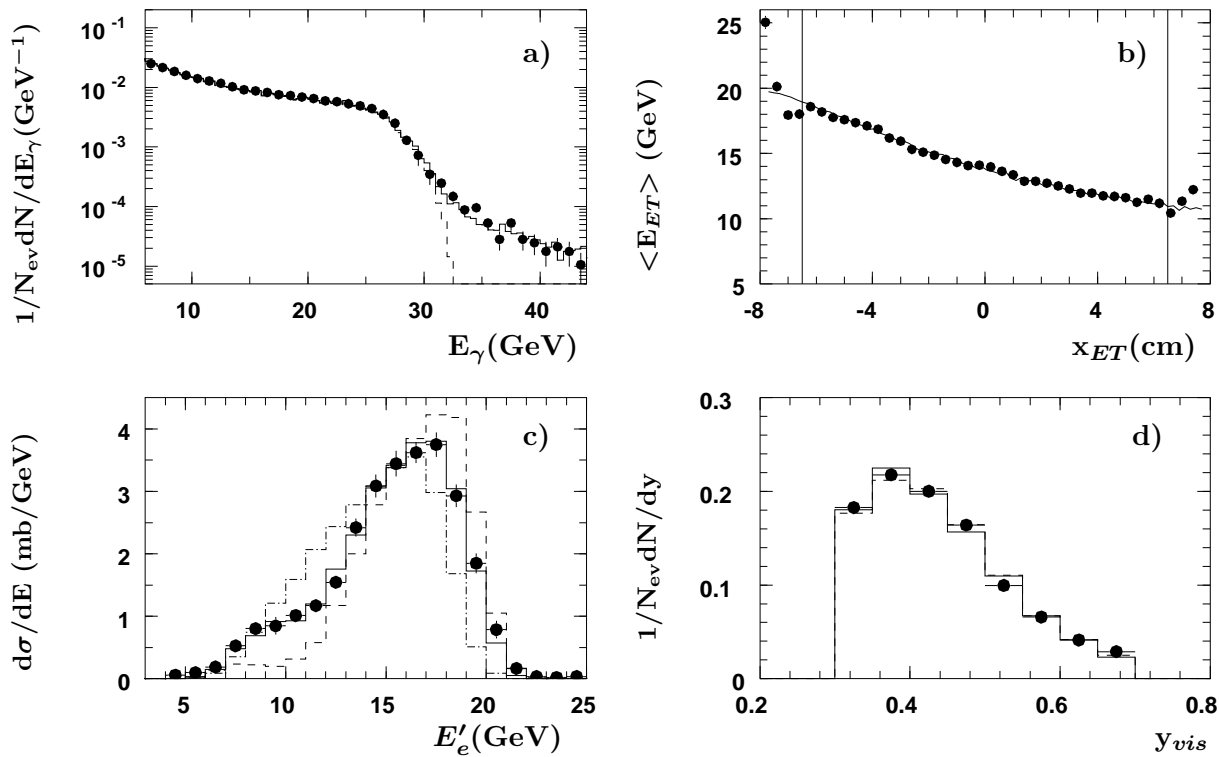}
\unboldmath
\end{picture}
\caption
  {Performance of the H1 luminosity system and
   the electron tagger acceptance.
   (a) Normalized photon energy distribution for Bethe-Heitler events
       in the data (symbols) and Monte Carlo with (full line) and
       without (dashed line) the event pile-up effect.
   (b) Correlation between the energy and the
       lateral coordinate of the impact point of scattered
       electrons in the electron tagger in the data (symbols)
       and Monte Carlo (full line).
       Vertical lines indicate the fiducial cut $|x_{ET}|<6.5$~cm
       used in the analysis.
   (c) Energy spectrum in the electron tagger for Bethe-Heitler events;
       data (symbols) are compared to  MC with measured
       $e$-beam tilt $\theta_x = -0.13$~mrad and  different offset
       values of the electron trajectory in the H1 interaction point:
       $x_{\rm off}$ = --0.5mm, +0.5mm, +1.5mm (dashed, full and
       dashed-dotted lines respectively).
   (d) $y$-distribution in the electron tagger for
       $\gamma{p}$ events in the range $0.3 < y < 0.7$ used in
       this analysis.
       The points represent the H1 data, histograms are Monte Carlo
       predictions for
       the models PYTHIA (solid) and PHOJET (dashed).}
\label{figLUMI}
\end{figure}

  The luminosity measurement utilizes the Bethe-Heitler~(BH)
  process \bh~\cite{BH}.
  Several methods for the measurement of the BH process
  can be exploited and they
  are described in detail in~\cite{QED}.
  In this analysis the
  method  based on the measurement of the photon energy
spectrum with $E_{\gamma} > E_{min} \simeq 8$~GeV is used.
  The value of $E_{min}$ is chosen such that it is well above the
  photon detector trigger threshold.
  The error in the luminosity measurement is then dominated by the
  precision of the energy calibration and by the correction for the
  complex structure of the proton bunches -- the so called
  ``satellite bunch" effect~\cite{QED}.
  After the final absolute energy calibration of the luminosity detectors
  the observed photon energy spectrum is shown in Fig.~\ref{figLUMI}a.
  It is well described by the BH process simulation taking
   into
  account energy resolution and pile-up effects (ie. several
  overlapping \bh~events in the same bunch crossing).
  The precision of the integrated luminosity measurement
  in different 1994 data
  samples varies between 1.5\% and 5.6\%.

  The electron tagger performs a double function. It both tags
  $\gamma{p}$~events
  and provides a measurement of the scaling variable $y$.
  Since the precision of the energy and coordinate
  reconstruction is not good enough in the areas close to the detector
  boundaries, $x_{ET} = \pm 7.7$~cm, a fiducial cut
  $|x_{ET}| < 6.5$~cm  is used in the analysis~(Fig.~\ref{figLUMI}b).
  A corresponding cut on $|y_{ET}|$ is redundant because of the
  confinement to the HERA bending plane.

  The acceptance $A(y,Q^2)$ of the electron tagger for scattered
  electrons depends strongly on the HERA electron beam optics,
  being most sensitive to the horizontal tilt $\theta_x$ and
  the horizontal offset $\Delta x_{\rm off}$
  with respect to the reference trajectory at the H1 interaction
  point.
  The electron beam tilt (typically 0.1~{\rm mrad})
  can be measured with a precision of $\pm 0.02$
  mrad by monitoring the position of the photon spot at the
  photon detector. The offset is not
  measured directly.
  The acceptance \AETBH~, integrated over $Q^2$,
can be determined using \bh~events.
  However, \AETBH~differs from the
  acceptance for photoproduction $A(y)$
  due to the different $Q^2$ dependences.

  The following procedure has been used to determine $A(y)$~for any
  data sample with constant beam conditions.
  First, the acceptance \AETBH~was measured from \bh~events.
  Then the Monte Carlo program simulating the  H1
  luminosity system together with the HERA beam optics
  was tuned to the data by varying
  $\Delta x_{\rm off}$, which is
  the only free parameter in the procedure.
  Fig.~\ref{figLUMI}c illustrates the sensitivity of the energy
  distribution in the electron tagger (and thus \AETBH) to the
  horizontal offset of the reference trajectory.
  A precision of $\Delta x_{\rm off} = \pm 0.2$~mm has been achieved
  by this procedure.
  Finally, the acceptance $A(y)$~was calculated using the measured tilt
  $\theta_x$ and the tuned value of $\Delta x_{\rm off}$.
  The errors have been estimated from calculations of \AETBH\ using
  extreme values of the parameters.
  A limited range of $0.3 < y < 0.7$ was used in the analysis,
  to avoid tails where the acceptance value is less than 20\%.
  Within this interval, errors between 3\% and 5\%
  were obtained in the value of $\int A(y)dy$
  for different data samples.
  We therefore conclude that 5\% can be used as a
  conservative estimate of the
  precision to which the electron tagger acceptance
  is known in this analysis.
  Fig.~\ref{figLUMI}d shows the comparison of the $y$ distributions
  in the data with Monte Carlo, using the two different models for
  photoproduction as described in section 3.

\section{Trigger Conditions and Event Selection}

  The data used for the measurement of $\stotgp$ were collected in
a short dedicated period during the 1994 data taking.
The HERA machine was operated with
153 colliding bunches of 27.6~GeV
positrons and 820~GeV protons.
In addition, 32 ``pilot" bunches, 17 proton and 15 positron, had no
counterpart and produced no $ep$ collisions enabling an estimate of
the beam induced background.
Two data samples were collected with different mean $z$-positions of the
$ep$ interaction vertex: the nominal position at $\bar{z}\,=\,4$~cm and a
position shifted in the proton direction at $\bar{z}\,=\,71$~cm
(see Fig.~\ref{figH1}).

 The data collected with the nominal vertex position correspond to
the integrated luminosity of $23.8\pm{0.4}$~nb$^{-1}$,
while the shifted vertex position data
correspond to  $23.8\pm{1.3}$~nb$^{-1}$.
The advantage of the shifted vertex data is the higher acceptance for
diffractive reactions in the region where diffractive processes can
be separated safely from the majority of non-diffractive events.

The data were taken with two independent trigger conditions.
   The first trigger condition, termed ``ToF-trigger", is formed by
the coincidence of a signal from the electron tagger
($E'_e > 4$~GeV) with a signal
from the ToF system coming within the time interval expected for
$ep$ interactions.
The ToF-trigger is fired by hadrons originating from  photon fragmentation.
This trigger is efficient for all classes
of photoproduction events including elastic vector meson production,
although it is affected by the BEMC material in front of the  ToF system.
The ToF trigger was enabled in both data samples.

 The second trigger condition, termed
``Ray-trigger", requires a coincidence of
the electron detector signal with at least one track
   pointing to the vertex region.
   The track condition is derived from the cylindrical MWPC and
requires a $p_T\;\gsim\;200$ \MeVa. This trigger
has been used in previous H1 analyses~\cite{stot_h1}. In the present
analysis the Ray-trigger is used for cross checks and
was activated during the run with nominal vertex position only.

 Both Ray- and ToF-triggers require in addition
the energy in the photon detector to be less than $2$~GeV.
This condition substantially reduces
the size of QED radiative corrections and
also suppresses accidental
coincidences of Bethe-Heitler events with beam induced background.

The triggered events are subjected to several offline cuts. The
fractional energy of the photon, as measured by the electron tagger,
is required to be in the interval $0.3 < y < 0.7$. The event vertex,
reconstructed from tracks in the CJC, must be within
$\pm\,30$~cm of the
mean $z$-position of the interaction point. The vertex $z$-position
distribution has a Gaussian shape with a sigma of $10$~cm reflecting the
length of the proton bunches. For ToF-triggered events
at least one reconstructed track in the
CJC or in the BPC is required in addition.
Similarly the Ray-triggered events are required to
have at least one CJC track in the region covered by the Ray-trigger.

Several sources of background contribute to the data samples. The main
background source in the ToF-triggered event samples is electron
interactions with residual gas (``beam-gas") or with material inside the
beam-pipe (``beam-wall"). This contribution is estimated using the data
from the non-colliding (pilot) electron bunches and amounts to $4\%$ and
$8\%$ respectively, in the event samples with shifted and non-shifted
vertex. For Ray-triggered events this background is negligible.
Another major source of background originates from
 accidental coincidences of the
electron tagger signal with events resulting from proton beam-gas
collisions within the nominal ep interaction region. This contribution
is estimated using special monitoring triggers, with looser triggering
conditions, and is about $3\%$ in the Ray-triggered data sample. For the
ToF-triggered events this background is negligible.
Still another type of background stems from the accidental coincidence
of electron beam induced background with a Bethe-Heitler process
induced signal in the electron tagger. This coincidence appears as
background when the associated photon escapes detection in the photon
detector, due to the small inefficiency of this detector. From the
measured rate of the Bethe-Heitler process this
background contribution is estimated to be about $1\%$ in all data samples.
Finally there is a small background contribution in all data samples
                                                 from the QED 2-photon
lepton pair production processes, with one photon emitted from the
incident electron and the other photon emitted from the proton. This
background was calculated using the LPAIR MC event
generator~\cite{lpair}
and  amounts to less than $0.2\%$ under the present trigger and
selection conditions.
\begin{figure}[htb]  \centering
  \boldmath
  \begin{picture}(160,120)
    \put( 65,110){\bf a)}
    \put(147,110){\bf b)}
    \put( 65,52){\bf c)}
    \put(147,52){\bf d)}
    \put( 152,7){$\eta$}
    \put( 55,7){$P_t$ (GeV/c)}
    \put( 50,67){$z - z_0 $ (cm)}
    \put( 150,67){$n_{ch}$}
    \put(  2,+85){\begin{sideways}{\small \bf events / 2 cm} \end{sideways}}
    \put( 83,+91){\begin{sideways}{\small \bf events} \end{sideways}}
    \put( 0,+15){\begin{sideways} $\frac{1}{N_{EV}}$
      {\small \bf (tracks / 0.1 GeV/c)} \end{sideways}}
    \put( 81,+21){\begin{sideways} $\frac{1}{N_{EV}}$
      {\small \bf (tracks / 0.1) } \end{sideways}}
  \epsfig
  {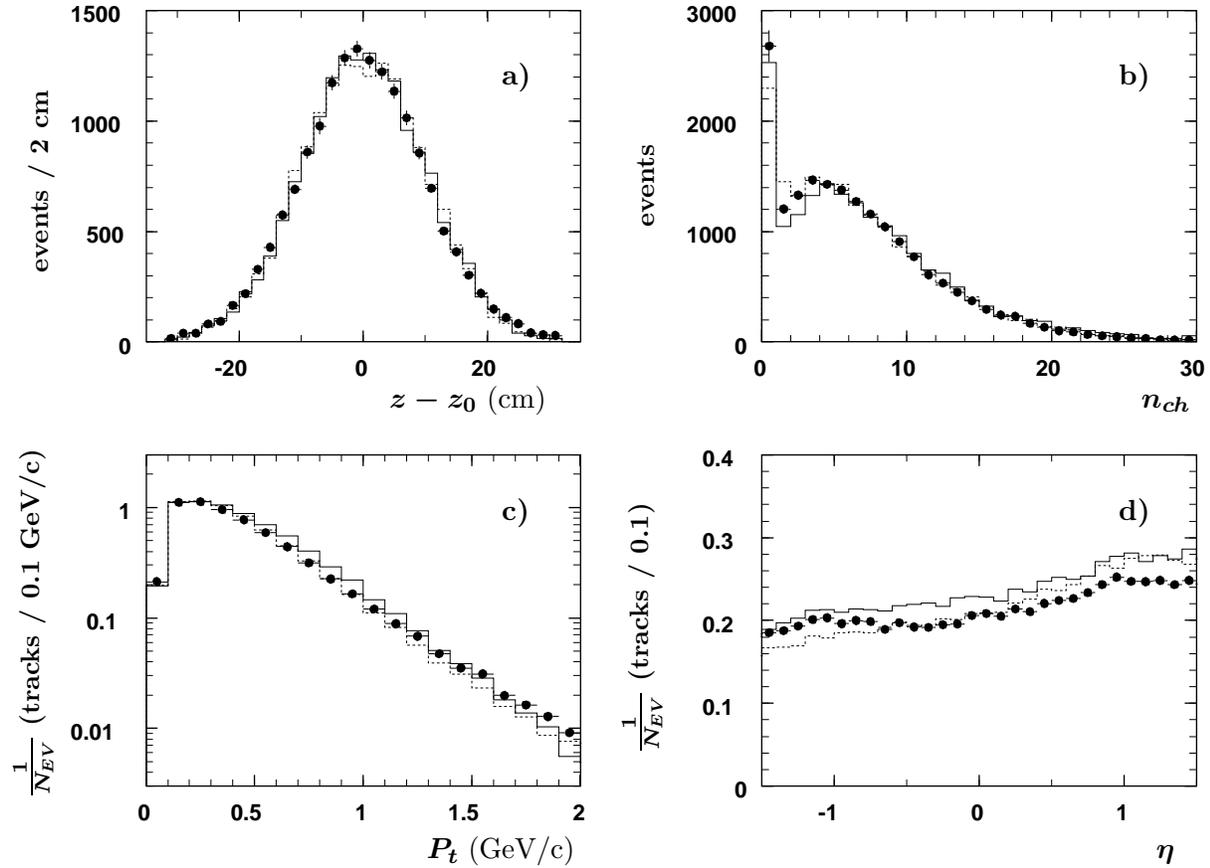}
\end{picture}
\unboldmath
\caption
{
The measured
$z$-position of the reconstructed event vertex~(a), multiplicity~(b),
$P_t$~(c) and $\eta$~(d) distributions of charged tracks in the data
(points) for the nominal vertex sample
compared with those in MC simulation using PHOJET (full histogram)
and PYTHIA (dotted histogram).
}
\label{figCP}
\end{figure}
The selected event samples vary between about 19,000 and 22,000 events. The
background is subtracted on a statistical basis in the analysis.

Fig.~\ref{figCP}
  shows a comparison between the nominal vertex ToF-triggered data
sample and the two MC simulations.
In Fig.~\ref{figCP}a the $z$-position of the
reconstructed event vertex is shown
and in Fig.~\ref{figCP}b-d the multiplicity, transverse
momentum and pseudo-rapidity of charged tracks are given.
Events with no CJC tracks (i.e. events which fulfil
the BPC requirement) do not contribute in the distributions of
Fig.~\ref{figCP}a,~c and d.
The agreement between data and simulations demonstrates that the MC
event generators reproduce well the main features seen in the data.
Especially the $z$-vertex distribution in
Fig.~\ref{figCP}a shows that the background
contamination in the data is small. Similar comparisons (not shown) for
the Ray-triggered event sample lead to the same conclusion.
%
%

%
\section{Cross Section Calculation Method}
%

 For each data sample the observed number of events~$N$ is related
to the differential $ep$ cross section~(\ref{eqWWA}) by the expression
\begin{equation}
  \frac{dN}{dydQ^2}={\cal L}\cdot\varepsilon(y)\cdot A(y,Q^2)
  \cdot\frac{d^2\sigma^{ep}}{dydQ^2},
  \label{eqS1}
\end{equation}
where ${\cal L}$ is the integrated luminosity, $\varepsilon(y)$ is
the efficiency of the trigger and selection criteria for the
main H1 detector, and $A(y,Q^2)$ is the acceptance of electron tagging
as described above. The $A(y,Q^2)$ is factorized out since it depends
only on the parameters of the scattered electron and not on the details
of the photoproduction process.
%
%
We assume
$\varepsilon(y)$ to be independent of $Q^2$ since the
 transverse momentum of the scattered electron is always small.
Integrating  equation~(\ref{eqS1})
over $y$ and $Q^2$ in the range from $y_{min}$ to $\,y_{max}$
and from $Q^2_{min}$ to $\,Q^2_{max}$ gives
\begin{equation}
  N=(1+\delta_{RC})\cdot{\cal L}\cdot F\cdot A\cdot\varepsilon\cdot\stotgp,
  \label{eqS2}
\end{equation}
using the photon flux integral $F$ and averaged values~\footnote
{
  A strict definition of the quantities averaged over $y$ and $Q^2$ is \\
  $\stotgp=\int\stotgp(ys)\varepsilon(y)A(y)F(y)dy/
  \int\varepsilon(y)A(y)F(y)dy,
  \,\,\,F=\int{F(y)dy},\,\,\, A=\int{A(y)F(y)dy/F}, \\
  \varepsilon=\int\varepsilon(y)A(y)F(y)dy/AF,$ with
  $F(y)=\int{F(y,Q^2)dQ^2}$ and
  $A(y)=\int\limits{A(y,Q^2)F(y,Q^2)dQ^2}/F(y)$}
of $A$, $\varepsilon$ and the cross section $\stotgp$.
An identical expression also holds for any
partial cross section $\sigI$ with only $\varepsilon$
depending on the sub-process {\it i}. Therefore, one has for their sum
\begin{equation}
  N=(1+\delta_{RC})\cdot{\cal L}\cdot F\cdot A\cdot
  \sum \varepsilon_i\sigI .
  \label{eqS3}
\end{equation}
Eq.(\ref{eqS3}) is also valid for the number of events
in any kinematic domain of the photoproduction process,
with $\varepsilon_i$ determined accordingly.
This enables us to find the
partial cross sections
by considering appropriate kinematic regions,
enriched by different sub-processes, and solving a
(generally over-constrained) system of equations for the
 $\sigI$, with efficiencies $\varepsilon_i$
calculated by MC simulations.

For our basic kinematic limits of
$y_{min}=0.3$~, $y_{max}=0.7$
 (corresponging to the range $165<W_{\gamma{p}}<252~\GeV$) and
$Q^2_{max} = 0.01~\GeV^2$
we find $F=0.0136$ and the acceptance $A$
varying from 0.546 to 0.570 depending on beam conditions.
This variation of $A$ is properly taken into account
in the analysis.
The MC efficiencies $\varepsilon_i$ for the
selected ToF-triggered samples are presented
in Table~\ref{tabEFF}~.
The average value of the radiative correction
$\delta_{RC}$ is estimated to be $(-1\pm{1})\%$ and $(+1\pm{1})\%$
for the ToF- and Ray-trigger samples, respectively.

%
%
%
\begin{table}[bth]
  \caption{ Efficiencies $\varepsilon_i$ (\%)
    for the different subprocesses of $\gamma p$ scattering
    as calculated using the Monte Carlo simulation
    based on the PHOJET(PYTHIA) models
    for various data samples.
    The $\eta_{max}<0$ and $\eta_{min}>1$ samples are used to find the
    diffractive contributions.}
  \label{tabEFF}
  \begin{center}
    \begin{tabular}{||c|c||c|c|c|c|c||} \hline\hline
      &&\multicolumn{5}{c||}{\em subprocess} \\ \cline{3-7}
      \raisebox{1.4ex}[2.ex][0ex]{\em sample} &
      \raisebox{1.4ex}[2.ex][0ex]{\em sub-sample} &
         {\em GD} & {\em PD} & {\em DD} & {\em EL} & {\em ND} \\
      \hline \hline
           & all events & 74(77) & 68(71) & 71(73) & 57(54) &  61(65) \\
      \raisebox{1.4ex}[2.ex][0ex]{shifted vertex} &
             $\eta_{max}<0$ & 31(28) & 28(18) & 15(17) & 52(50) & 0.1(0.1) \\
      \raisebox{1.4ex}[2.ex][0ex]{ToF trigger} &
    $\eta_{min}>1$ & 0.2(0.3) & 21(26) & 3.4(3.1) & 0.4(0.4) & 0.1(0.1) \\
      \hline\hline
      nominal vertex & & & & & & \\
      ToF trigger &
               \raisebox{1.4ex}[2.ex][0ex]{all events} &
               \raisebox{1.4ex}[2.ex][0ex]{66(73)} &
               \raisebox{1.4ex}[2.ex][0ex]{42(46)} &
               \raisebox{1.4ex}[2.ex][0ex]{68(72)} &
               \raisebox{1.4ex}[2.ex][0ex]{29(26)} &
               \raisebox{1.4ex}[2.ex][0ex]{65(70)} \\
      \hline\hline
      nominal vertex & & & & & & \\
      Ray trigger &
               \raisebox{1.4ex}[2.ex][0ex]{all events} &
               \raisebox{1.4ex}[2.ex][0ex]{57(65)} &
               \raisebox{1.4ex}[2.ex][0ex]{8(14)} &
               \raisebox{1.4ex}[2.ex][0ex]{62(52)} &
               \raisebox{1.4ex}[2.ex][0ex]{0(0)} &
               \raisebox{1.4ex}[2.ex][0ex]{95(94)} \\
      \hline\hline
    \end{tabular}
  \end{center}
\end{table}

\section{Cross Section Measurement}

To measure the total $\gamma{p}$ cross section we
use data with the nominal vertex position, where
the uncertainty in the luminosity calculation is significantly
smaller than in the data taken with the shifted vertex position.
However, as can be seen from Table~\ref{tabEFF} the efficiencies for
the diffractive channels are higher in the shifted vertex data sample.
Therefore,
for the determination of the diffractive contributions to the $\gamma{p}$
cross section we use the data taken with shifted $z$-vertex position.
The analysis and the cross checks
 are described in the following subsections $8.1-8.3$.

\subsection{Diffractive Contributions}

To measure the diffractive $\gamma{p}$ cross sections
we choose the variables in which a separation of
diffractive and non-diffractive contributions is least model dependent.
These variables, $\eta_{max}$ and $\eta_{min}$,
are related to the central rapidity gap in the hadronic final
state.

For each event $\eta_{max}$  is defined as the
maximum pseudo-rapidity of all reconstructed charged tracks and
all clusters in the LAr calorimeter
with energy larger than $400~\MeV$.
In diffractive events $\eta_{max}$ indicates the maximum
pseudo-rapidity of secondary hadrons from
photon dissociation. It was shown in a recent H1 analysis of
diffractive photoproduction~\cite{h1_diff} that
%
%
the spectrum of $\eta_{max}$ for non-diffractive
events falls nearly exponentially with decreasing $\eta_{max}$, whilst the
rate of photon diffractive dissociation
depends only weakly on this variable. Events with elastic
vector meson production have the largest possible width of the rapidity gap
and are concentrated at lower values of the $\eta_{max}$ spectrum.
%
%
%
%
\begin{figure}[htb]  \centering
  \boldmath
  \begin{picture}(160,100)
%
    \put( 150,6){$\eta_{min}$}
    \put( 65,6){$\eta_{max}$}
    \put( 3,+70){\begin{sideways}{\small \bf events / 0.2} \end{sideways}}
    \put( 83,+70){\begin{sideways}{\small \bf events / 0.4} \end{sideways}}
    \put(  65,85){\small \bf (a)}
    \put( 145,85){\small \bf (b)}
    \epsfig
    {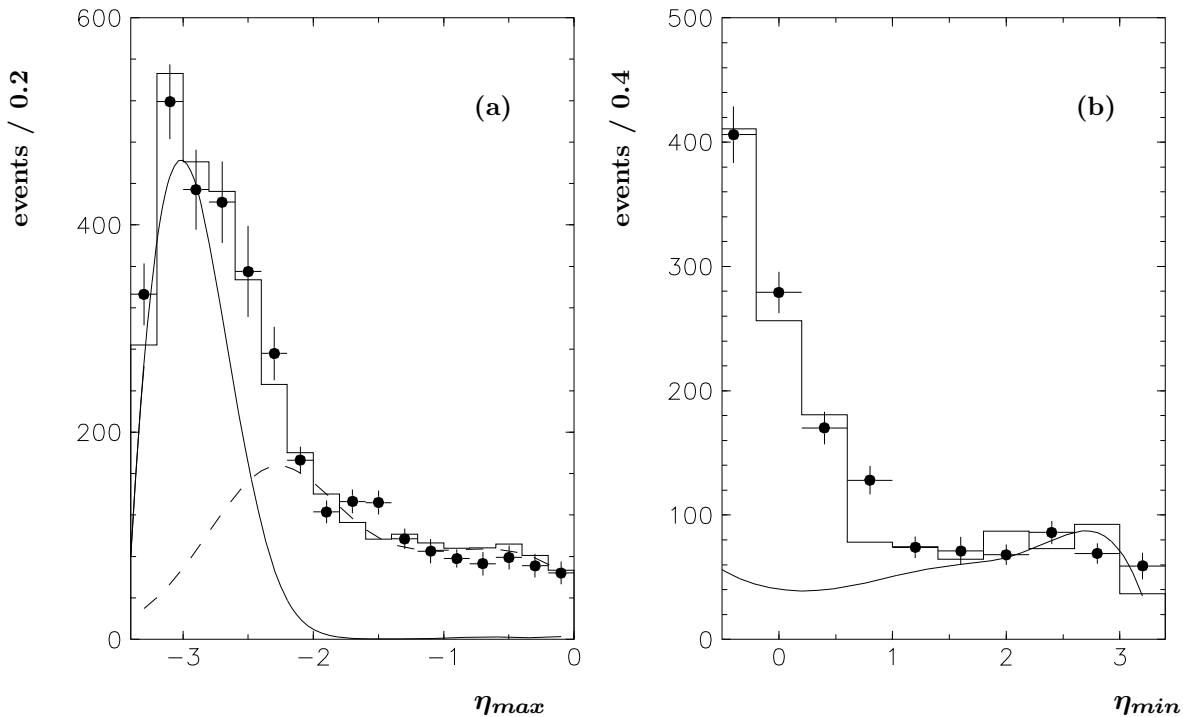}
  \end{picture}
  \unboldmath
 \caption
 {
   The $\eta_{max}$ (a) and $\eta_{min}$ (b) distributions
   for the sample with shifted vertex. The data are
   shown with solid circles.
   The histogram represents the result of a combined fit
   based on the PHOJET model
   summing all five contributions (EL, PD, GD, DD and ND)
   and using $\sigma_{DD}=20~\mu$b.
   The full curve in (a) represents the sum of elastic and single proton
   dissociation. The  sum of single photon and double dissociation
   contributions in (a) is represented by the dashed curve. The sum of
   single proton and double dissociation in (b) is shown by the full curve.
   }
 \normalsize
 \label{figETAM}
\end{figure}

The $\eta_{max}$ distribution for the data sample with
shifted interaction vertex is shown in Fig.~\ref{figETAM}a, where
we additionally
require the calorimetric energy with $\eta\,>\,1$ to be less than
$1~\GeV$. This requirement reduces the contribution from non-diffractive
events.
 MC calculations show that
all four diffractive reactions contribute to the $\eta_{max}$
spectrum in the range $-3.5<\eta_{max}<0$,
 whilst the contribution from non-diffractive processes is
negligible.
Below $\eta_{max}=-2$ the spectrum is dominated by
elastic and single proton dissociation channels, where
$\eta_{max}$ is determined by the maximum pseudo-rapidity
of the vector meson decay products.
These two contributions  are practically indistinguishable by
shape, but have different efficiencies. For proton diffractive
dissociation only the events with a low mass proton system $(M<10~\GeV/c^2$),
where secondary particles escape detection very close to the proton
beam direction, contribute to the $\eta_{max}$ spectrum.
 The region $-2<\eta_{max}<0$ is dominated
by single photon and double diffractive
dissociation contributions, which have again
a similar shape, but different acceptances.

The $\eta_{min}$ variable is defined as the minimum pseudo-rapidity in the
interval $-2<\eta<3.5$, of all charged tracks and of all calorimeter
clusters with an energy larger than $400~\MeV$. An additional condition is
that the event has a reconstructed charged track with $\eta<-2.4$.
The latter condition is necessary
for tagging a low mass hadronic system on the photon side.
In diffractive events $\eta_{min}$ indicates
the minimum pseudo-rapidity of secondary hadrons from
proton dissociation.
The measured $\eta_{min}$ distribution
for the data sample with shifted interaction vertex
is shown in Fig.~\ref{figETAM}b
and is dominated by the non-diffractive contribution,
which falls nearly exponentially
with increasing $\eta_{min}$. However, there is a flat part of the spectrum
with $\eta_{min}>1$ caused by proton
diffraction dissociation processes,
where the mass of the hadronic system produced on the proton side is
larger than about $5~\GeV/c^2$.
Both single and double dissociation contributions have a similar shape.
The acceptances for various partial processes to contribute to the
$\eta_{max}$ and $\eta_{min}$ distributions are presented in
Table~\ref{tabEFF}, calculated using the PHOJET and PYTHIA MC models.

\begin{figure}[htb]  \centering
   \boldmath
   \begin{picture}(170,94)(0,6)
     \put( 130,8){$\sigma_{DD}\,$ ($\mu${\bf b})}
     \put( 55,8){$\sigma_{DD}\,$ ($\mu${\bf b})}
     \put( 3,+79){\begin{sideways}$\sigma_i\,$ ($\mu${\bf b})\end{sideways}}
     \put( 83,+79){\begin{sideways}$\sigma_i\,$ ($\mu${\bf b})\end{sideways}}
     \epsfig
     {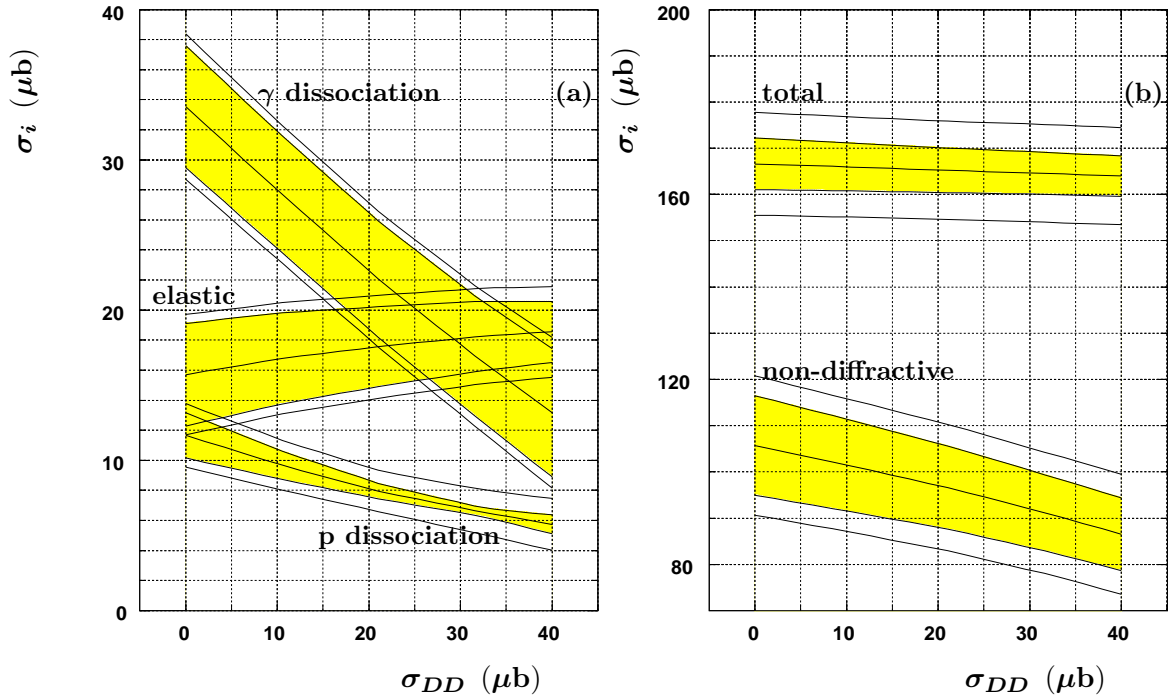}
     \put(-124,86){\small \bf $\gamma$ dissociation}
     \put(-138,59){\small \bf elastic}
     \put(-116,27){\small \bf p dissociation}
     \put(-57,86){\small \bf total}
     \put(-57,49){\small \bf non-diffractive}
     \put(-84.5,86){\small \bf (a)}
     \put( -9.0,86){\small \bf (b)}
   \end{picture}
   \unboldmath
   \caption
   {
     The measured partial diffractive~(a)
     and non-diffractive and total~(b) $\gamma p$  cross sections
     as a function of the assumed double dissociation cross section.
     The errors are shown as bands of $\pm$1 standard deviation.
     The wide bands correspond to full statistical and systematic
     errors added in quadrature. The narrow grey bands show
     the systematic errors due to the  model dependence
     described in the text.
     }
   \label{figSIGMAI}
 \end{figure}

To obtain the diffractive $\gamma{p}$ cross sections we make a
combined fit of the $\eta_{max}$ and $\eta_{min}$ distributions in the
intervals $-3.5<\eta_{max}<0$ and $1<\eta_{min}<3.5$
using formula~(\ref{eqS3}).
In this fit procedure the shapes of the spectra for each
partial contribution are fixed from the MC calculations, while the
cross sections $\sigma_i$ for single dissociation and elastic
reactions are left as free fit parameters.
Since only three of the four diffractive cross sections can be reliably
extracted from the fit we make an additional
assumption about the value of the double dissociation cross section
varying it from 0 to 40~$\mu$b.
The upper limit is chosen to be about two times larger
than the value expected from the low energy measurements
extrapolated using Regge-type formalism~\cite{goulianos_conf}.
Our attempts to determine
the double dissociation contribution directly from the data
by observing a high mass dissociation of both the proton and the photon
 give results within this interval,
but are inconclusive.
In order to estimate the contribution of the tail of the
non-diffractive reaction into the fitted $\eta_{min}$ region,
the non-diffractive cross section was fixed
in the fit to describe the part
of the $\eta_{min}$ spectrum below $\eta_{min}=1$.
An example of a fit using the PHOJET model and the assumption
$\sigma_{DD}=20~\mu$b is shown in Fig.\ref{figETAM}.
The fit describes the data well.
The cross sections $\sigma_{EL}$, $\sigma_{GD}$ and
$\sigma_{PD}$ obtained are displayed in Fig.\ref{figSIGMAI}a as
functions of $\sigma_{DD}$.
The errors are dominated by systematic uncertainties
due to model dependence and are shown by grey bands.

%
%
%

%
The model dependence was studied by using different MC generators
(PYTHIA and PHOJET) and, in addition, by varying the main
parameters of the diffractive model within the MC generator. These
parameters are
the value of the nuclear slope~$B$,
the minimum value for the mass of the dissociated system
and the form of the mass dependence of the cross section:
\begin{itemize}
  \item
    The value of the {\em nuclear slope} was varied by
    $\Delta{B}=\pm\,4~(\GeVa)^{-2}$
    in the elastic process and half this range
    in the single dissociation processes. This
    represents a conservative estimate of the uncertainty in
    the extrapolations
    of the measured slope from lower energies~\cite{chapin}.
  \item
    The uncertainty in the description of the measured {\em low mass
      part} of the dissociated mass spectrum~\cite{chapin,goulianos}
    was
    conservatively estimated by increasing the value of the
    minimum mass of the diffractively produced hadronic system
    by $0.2~\GeV/c^2$.
  \item
    {\em Mass dependence }
    of single and double diffractive dissociation.
    The photoproduction data at lower energy~\cite{chapin} and
    hadron hadron diffractive dissociation at
    $\sqrt{s}=546~\GeV$~\cite{ua4} are well described by a
    phenomenological $1/M^2$ dependence.
    This mass dependence
    is implemented in the MC models used for the analysis.
    However, the predictions of Regge theory with a
    supercritical pomeron trajectory give after an
    integration over $t$ $1/M^{2\alpha{(0)}}$,
    where $\alpha{(0)}$ is the value of the
    intercept of the pomeron trajectory at $t=0$.
    The phenomenological fit of the total
    cross section~\cite{stot_ld} gives $\alpha{(0)}=1.08$.
    In order to estimate this part of the model dependence we set
    the mass distribution also to $1/M^{2.2}$ in the MC calculations.
\end{itemize}

For every variation of a parameter we average the results
and ascribe half of the spread to be the
corresponding systematic error. This is then added to other errors
in quadrature. Similarly, we average the results
obtained with the PYTHIA and PHOJET models.
The final results are shown
in Fig.\ref{figSIGMAI}a.
     The wider bands correspond to full statistical and systematic
     errors added in quadrature. The narrower grey bands show
     the contribution to
     the systematic errors from the model dependence
     described above.
The main sources of
the model independent contribution to the
 systematic errors are the uncertainties in
the luminosity measurement ($5.6\%$),
in the acceptance of the electron tagger ($5.0\%$,
which affect all the results)
and an uncertainty in the statistical background subtraction
affecting only the  elastic channel~($8\%$).
The statistical errors are much smaller than the
systematic errors in all cases except for proton dissociation where
they are comparable.
Variation of the non-diffractive contribution to the fitted distributions
by a factor of two, changing the fit interval and variation
of the different requirements used for the data selection alter the
results only within the statistical errors.

One can see from Fig.\ref{figSIGMAI}a that the elastic cross
section is almost independent of any assumption made about $\sigma_{DD}$.
Proton dissociation and especially photon dissociation
show a stronger correlation with the assumed value of $\sigma_{DD}$.
However, at any value of $\sigma_{DD}$,
the single photon dissociation is substantially larger than
the single proton dissociation, in contrast to the assumption
that they are equal, made in earlier
$\stotgp$ analyses \cite{stot_h1,stot_zeus} at HERA.

The values of the diffractive cross sections averaged over
$\sigma_{DD}$, are presented in Table~\ref{tabRESULTS}.

%
%
%
\begin{table}[bth]
  \caption{ Results for partial and total
    $\gamma p$ cross sections under the assumption that the double
    dissociation cross section is in the range
    $0<\,\sigma_{DD}<\,40~\mu$b.
    The first error is statistical, the second one is systematic,
    and the third error reflects the
    systematic uncertainty due to
    the assumption on $\sigma_{DD}$ . Their sum in quadrature
    is given as the full error.
    Note, that due to the error correlation in the partial cross sections
    the error of the total cross section is relatively smaller.}
  \label{tabRESULTS}
  \begin{center}
    \begin{tabular}{||c||rcrcrcr|crc||} \hline\hline
      {\em process} &
      \multicolumn{7}{c|}{{\em cross section} ($\mu$b)} &
      \multicolumn{3}{c||}{{\em full error} ($\mu$b)} \\
      \hline\hline
      $\sigma(\gamma{p}\rightarrow\,XY)$,\,  {\em DD} &
       \multicolumn{7}{c|}{20 $\pm$ 20 (assumed)} & $\quad$
        & --- &  \\
      $\sigma(\gamma{p}\rightarrow\,Xp)$,\, {\em GD} &
       23.4 & $\pm$ & 2.6 & $\pm$ & 4.3  & $\pm$ & 10.2 && 11.3 & \\
      $\sigma(\gamma{p}\rightarrow\,VY)$,\,  {\em PD} &
       8.7 & $\pm$ & 1.5 & $\pm$ & 1.5  & $\pm$ &  3.0 &&  3.6 & \\
      $\sigma(\gamma{p}\rightarrow\,Vp)$,\, {\em EL} &
      17.1 & $\pm$ & 1.6 & $\pm$ & 3.7  & $\pm$ &  1.4 &&  4.3 & \\
      \hline
      {\em EL + GD + PD + DD } &
      69.2 & $\pm$ & 3.4 & $\pm$ & 8.8 & $\pm$ &  9.3 && 13.2 & \\
             {\em ND} &
      96.1 & $\pm$ & 3.5 & $\pm$ & 14.7 & $\pm$ &  9.6 && 17.9 & \\
      \hline
      {\em Total} &
                165.3 & $\pm$ & 2.3 & $\pm$ & 10.9 & $\pm$ &  1.3 && 11.2 & \\
      \hline\hline
    \end{tabular}
  \end{center}
\end{table}
%
%

\subsection{Non-Diffractive and Total Cross Sections}

The  measurement of the total $\gamma{p}$ and non-diffractive
cross sections is based on the ToF-triggered sample with the
nominal vertex.
The efficiencies for the different subprocesses are given
in Table~\ref{tabEFF}.
As mentioned above, the efficiencies for elastic and
proton diffraction channels using nominal vertex
are about half those for shifted vertex data making the
determination of diffractive contributions from
the nominal vertex sample less reliable. The
diffractive cross sections are therefore assumed to be those
found from the shifted vertex
data, as described in the previous section, and
the nominal vertex data are used to determine only the missing
non-diffractive cross section.
This was done by solving eq.(\ref{eqS3}) for $\sigma_{ND}$, then calculating
$\stotgp$
as a sum over five partial contributions, all errors being
properly propagated including a correlation between diffractive and
non-diffractive cross sections.

The $\sigma_{ND}$ and $\stotgp$ obtained as a function of
the assumed value of $\sigma_{DD}$ are shown in Fig.\ref{figSIGMAI}b.
The model dependence displayed has been
studied in exactly the same way as in the
analysis above. The model uncertainty of $\stotgp$ and $\sigma_{ND}$ is
dominated by the difference between
PHOJET and PYTHIA models which enters the calculations
via the different efficiencies for the  ND, GD and DD channels
shown in Table~\ref{tabEFF}.

The total cross section is remarkably insensitive
to assumption about $\sigma_{DD}$ and  changes only by 2.6~$\mu$b
as $\sigma_{DD}$ varies between 0 and 40~$\mu$b.
This is because the efficiencies $\varepsilon_i$
for GD, DD and  ND are very similar
(see Table~\ref{tabEFF}). The sum
$\sigGD + \sigDD + \sigND$ is therefore practically fixed by eq.(\ref{eqS3}),
whilst the total contribution of $\sigma_{PD}$ and $\sigma_{EL}$
in eq.(\ref{eqS3}) is weakly dependent on $\sigma_{DD}$.

The results, averaged over $\sigma_{DD}=$0--40~$\mu$b are
given in Table~\ref{tabRESULTS} along with the diffractive
cross sections.
We finally obtain the total photoproduction cross section
for an average $W_{\gamma{p}}$ of $200~\GeV$
\begin{center}
$\stotgp$ = 165.3 $\pm$ 2.3(stat.) $\pm$ 10.9(syst.) $\mu$b. \\
\end{center}

The statistical error reflects all the relevant
statistical uncertainties for the shifted and nominal vertex
data samples, as well as those from the MC calculations.
The various contributions to the systematic errors are listed
in Table~\ref{tabERRORS}. The dominant sources
are the uncertainty of the e-tagger acceptance ($\pm 8.5 \mu$b)
and the difference between the PYTHIA and PHOJET models
($\pm 5.1 \mu$b).

%
%
%
\begin{table}[htp]
  \caption
    {Different contributions to the systematic error of the $\stotgp$
    measurement.}
  \label{tabERRORS}
  \begin{center}
    \begin{tabular}{||l|c|c||}
      \hline\hline
      &\multicolumn{2}{c||}{{\em Error} ($\mu$b)} \\ \cline{2-3}
      \raisebox{1.4ex}[2.ex][0ex]{\em Source} &
      {\em ToF-trigger} & {\em Ray-trigger} \\
      \hline\hline
      {assumption of $\sigma_{DD}\,<\,40\mu$b}   & 1.3 &  5.4  \\
      {efficiency uncertainty due to model dependence}
                                                     & 5.6 & 5.7 \\
      {syst. errors of diffr. cross sections (without model dependence)}
                                                     & 1.3 & 4.2 \\
\hline
      {1.6\% uncertainty in the luminosity measurement}  &
            \multicolumn{2}{c||}{2.8} \\
      {0.4\% uncertainty in the fraction of luminosity in satellite bunches}
            & \multicolumn{2}{c||}{0.5} \\
      {5\% uncertainty in the $e$-tagger acceptance}       &
              \multicolumn{2}{c||}{8.5} \\
      {1\% uncertainty in QED radiative corrections} &
           \multicolumn{2}{c||}{1.7} \\
      \hline
      {\em Total}                             &10.9 & 12.7 \\
      \hline\hline
    \end{tabular}
  \end{center}
\end{table}

%
\subsection{Cross Checks of the Total Cross Section Measurement}

The data taken with the Ray-trigger allow
a cross check of the  model-dependent acceptances
$\varepsilon_i$. Since ToF is sensitive only to particles
with $\eta < -2$ and the Ray trigger is fired by particles
around $\eta = 0$, the two triggers are quite independent
of each other. We have compared the fraction of ToF-triggered
events in the Ray-triggered data sample
with predictions from the MC simulations. This fraction
is $70\%$ in the data, which is $2\%$ higher than the fraction
calculated with the PHOJET model and $3\%$ lower than that predicted
by the PYTHIA model.
This cross check suggests that the true value
of the acceptance lies between the PYTHIA and PHOJET
estimates, and validates our procedure of averaging
the cross sections over the two models.

A second determination of the cross section has been made
using the Ray-trigger data sample, giving $\stotgp=162\pm{2}\pm{13}~\mu$b.
Here the relative contributions of the subprocesses are taken from
the above measurements (section~8.1)
for different values of \sigDD.
  This second determination of $\stotgp$ is consistent with the
first result described above.
The larger systematic error of this result reflects the larger
difference in the efficiency for different reactions in the Ray-trigger
data sample, compared to those in the sample taken with the ToF-trigger.
  The Ray-trigger selection has
a higher acceptance for the non-diffractive reactions,
but shows a nearly vanishing efficiency for
the elastic and the single proton dissociation processes.
  In addition, since it depends strongly on the transverse
momentum and multiplicity of
charged particles,
the Ray-trigger has a higher sensitivity
to the details of the hadronic final state simulation.
The contributions to the systematic error of this measurement
are also given in Table~\ref{tabERRORS}.

A determination using only the shifted vertex data resulted in
$\stotgp=166\pm{2}\pm{15}~\mu$b.
The larger systematic error is a result of the larger uncertainty
of the luminosity measurement.

%
%
%
%
%
\begin{figure}[htb]  \centering
  \boldmath
  \begin{picture}(140,110)
%
    \put( 54,90){\small \bf H1 , this analysis   }
    \put( 54,82.5){\small \bf ZEUS }
    \put( 54,75){\small \bf low-energy data }
    \put(  75,  5){$W_{\gamma{p}}\,$ ({\bf GeV})}
    \put( 3,+75){\begin{sideways}$\stotgp\,$ ($\mu${\bf b})\end{sideways}}
    \epsfig
    {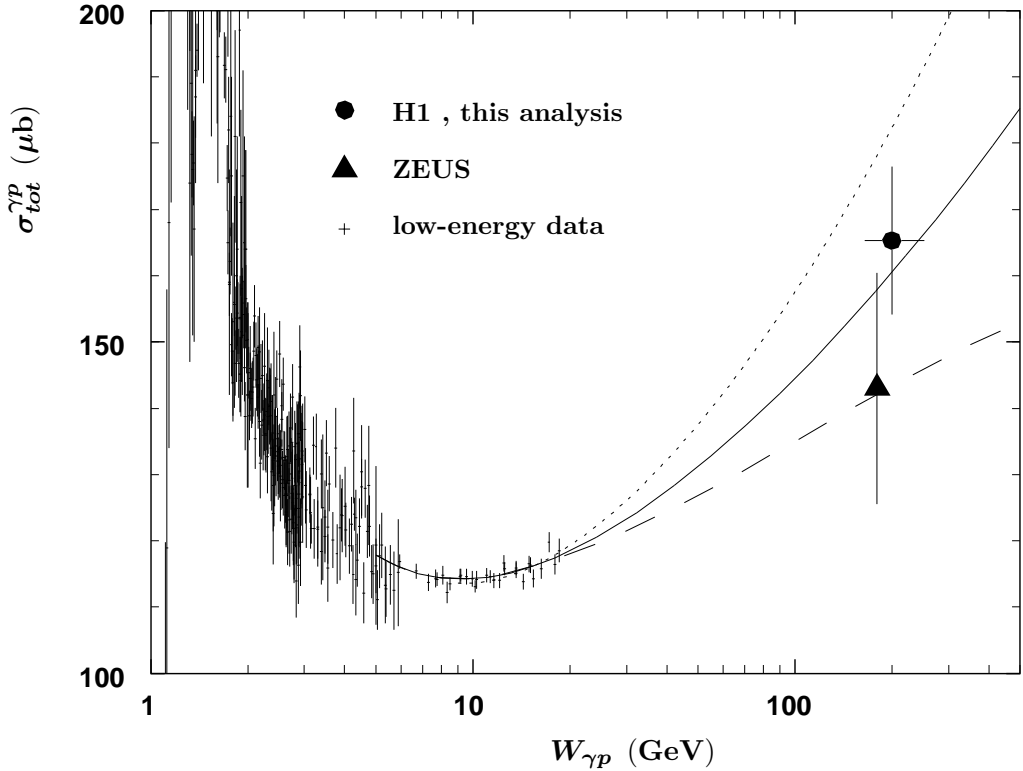}
  \end{picture}
\unboldmath
\caption
{
  Total photoproduction cross section as a function of the $\gamma{p}$
  centre of mass energy $W_{\gamma{p}}$. The solid line is the prediction of
  the DL~[4]
 combined fit of hadron hadron and low energy photoproduction data and
  the dashed line is the ALLM~[5]
  parameterization. The dotted line presents
  the DL parameterization obtained after the recent measurement of the total
  $p\bar{p}$ cross section by CDF~[8].
  }
\label{figSTOT}
\end{figure}
%
%
%

%
%
%
%
%
%
%
%

\section{Discussion}

The energy dependence of $\stotgp$ is shown in Fig.~\ref{figSTOT}
with the low energy data~\cite{stot_lowe}
and the present measurement together with a
recent result from ZEUS~\cite{stot_zeus}.
The data are compared with predictions made by
A.Donnachie and P.V.Landshoff~(DL)~\cite{stot_ld} and
H.Abramovich, E.M.Levin, A.Levy and U.Maor~(ALLM)~\cite{stot_allm}.
The DL curve presents a parameterization of a
universal rise of the cross section in hadron hadron and
low energy photon hadron
collisions. In this parameterization the high energy cross section
behaviour is described by a function $W_{\gamma{p}}^{2\Delta}$ with
$\Delta = 0.0808$.
The dotted line in Fig.~\ref{figSTOT} represents a further DL-type
  parameterization,
which takes into account the recent CDF measurement of the total $p\bar{p}$
cross section~\cite{stot_cdf} with $\Delta = 0.11$.
The ALLM is a Regge-type cross section parameterization
for real and virtual photon proton collisions with $\Delta = 0.045$.

In Fig.~\ref{figSIGMAI} the measurement
of the diffractive  cross
section contributions is shown
as a function of the assumed value for the double
diffractive dissociation $\gamma{p}$ cross section.
One should note that the results presented here on photon and proton
diffractive
dissociation cross sections are the first measurements of these
quantities at HERA energies.
This measurement can be compared with
predictions of A.Capella, A.Kaidalov, C.Merino and J.Tran Thanh Van
(CKMT)~\cite{diff_cap},
G.A.Schuler and T.Sj\"ostrand (SaS)~\cite{diff_ss},
and E.Gotsman, E.M.Levin and U.Maor
(GLM)~\cite{stot_glm}\footnote{The published results of the GLM calculation
are scaled to a dissociation mass interval of $M^2<0.1W_{\gamma{p}}^2$.}.
These models are based on different assumptions about the structure and
dissociation of the  photon and the proton.
{}From these models only the CKMT predictions are based on theoretical
calculations
using the Regge model, taking into account absorptive corrections for
all the diffractive reactions measured here.
The comparison of these models with the data is shown
in Table~\ref{tabMODELS}, where the data are presented for a fixed
value of the double dissociation cross section~($\sigma_{DD}=15\mu$b).
This particular choice is made according to the predictions of
the models.
%
%
%
%
%
%
\begin{table}[bth]
  \caption {The comparison of the diffractive cross section
    calculations
    with the H1 measurement. The data are
    presented for a fixed value $\sigma_{DD}=15\mu$b.
    }
  \label{tabMODELS}
  \begin{center}
    \begin{tabular}[h]{|c||c|c|c|c|}
      \hline
      &\multicolumn{4}{c|}{{\em Cross Sections}~($\mu$b)} \\ \cline{2-5}
      \raisebox{1.4ex}[2.ex][0ex]{\em Reaction} &
      Data & CKMT & SaS & GLM \\
      \hline\hline
      $\sigma(\gamma{p}\rightarrow\,Vp)$,\, {\em EL} &
      $17\pm\,4$ & 17 & 16 & 17 \\
      \hline
      $\sigma(\gamma{p}\rightarrow\,Xp)$,\, {\em GD} &
      $26\pm\,5$ & 25 & 13 & 18 \\
      \hline
      $\sigma(\gamma{p}\rightarrow\,VY)$,\,  {\em PD} &
      $9\pm\,2$ & 7 & 10 & 15 \\
      \hline
      $\sigma(\gamma{p}\rightarrow\,XY)$,\,  {\em DD} &
      $ 15 $ & 15 & 13 & 15 \\
      \hline
    \end{tabular}
  \end{center}
\end{table}
The predictions for the elastic reaction are in good agreement  with
the H1 measurement as well as with the recent result from
the ZEUS collaboration
($18\pm\,7\,\mu$b)~\cite{stot_zeus}.
However, the relative contribution of single proton dissociation
is observed to be about three times lower than that of  photon
dissociation. This observation disagrees with the predictions of the SaS
and GLM models. In the latter the ratio of the proton to the photon
single diffractive dissociation is obtained using
the quark counting rule.
The results of the CKMT calculations are
supported by the present measurement.

\section{Conclusion}
Using the H1 detector at HERA results on
$\gamma p$ scattering at the average c.m. energy of
$W_{\gamma{p}}=200~\GeV$ are obtained.
The total photoproduction cross section is measured to be
$\stotgp=165\pm{2}~(stat.)\pm{11}~(syst.)~\mu$b
replacing our previous result~\cite{stot_h1}.
The extracted diffractive and non-diffractive
contributions to the cross section are
presented as a function of the assumed value for the double diffractive
dissociation cross section $\sigDD$~.
The cross section of single photon diffractive
dissociation is observed to be
substantially higher than the single proton diffractive
dissociation cross section.
The elastic cross section depends only weakly on the assumption
made on $\sigDD$ and is found to be $17.1\pm 4.3~\mu$b~.

\par\noindent
{\bf Acknowledgments}

\noindent
We are grateful to the HERA machine group whose
outstanding efforts made this experiment possible. We appreciate the
immense effort of the engineers and technicians who constructed and
maintain the detector. We thank the funding agencies for
their financial support of the experiment. We wish to thank the DESY
directorate for the hospitality extended to the non-DESY members
of the collaboration. We are indebted to R.Engel and T.Sj\"ostrand
for help concerning the MC event generators used in this
analysis.
The stimulating discussions with M.Ryskin and
A.Kaidalov are gratefully acknowledged.

\renewcommand{\baselinestretch}{1.0}
{\Large\normalsize}

\end{document}